\documentclass[aps,apl]{revtex4}

\usepackage{amsmath,amssymb}
\input epsf
\usepackage{verbatim}
\usepackage{graphicx}
\usepackage{bm}
\usepackage{natbib}
\def\pmb#1{\setbox0=\hbox{#1}
\kern-.025em\copy0\kern-\wd0 \kern-.05em\copy0\kern-\wd0
\kern-.025em\raise.0433em\box0}

\newcommand{\C}{\mathbb C}

\newcommand{\beq}{\begin{equation}}
\newcommand{\eeq}{\end{equation}}
\newcommand{\ba}{\begin{eqnarray}}
\newcommand{\ea}{\end{eqnarray}}

\input epsf

\usepackage[english]{babel}
\begin{document}
%\doublespace

\title[]{Non-Euclidean elastodynamic cloaking theory and application to control of surface seismic waves with pillars atop a thick plate
% lying atop a
 % Faqir's bed
 % of nails
}
\author{Andr\'e Diatta$^{1}$,Younes Achaoui$^{1}$ and S\'ebastien Guenneau$^{1}$}
\affiliation{1.  Aix$-$Marseille Univ., CNRS, Centrale Marseille, Institut Fresnel, 13013 Marseille, France
}

\begin{abstract}
Recent experiments on control of anthropic seismic sources in a soil structured with boreholes [Phys. Rev. Lett. 112, 133901 (2014)] and on filtering effects by a forest of trees via analysis of seismic noise [Scientific Reports 6, 19238 (2016)] have fueled interest in large scale phononic crystals and mechanical metamaterials, also known as seismic metamaterials. The downside of these two inspirational works is the rather high frequency range of observed interactions between surface Rayleigh waves and structured soil compared with frequencies of interest for civil engineers, but in [AIP Advances 6, 121707 (2016)], a soil structured with concrete columns distributed within two specially designed seismic cloaks thanks to a combination of transformational elastodynamics and effective medium theory was shown to detour Rayleigh waves of frequencies lower than $10$ Hz around a cylindrical region. The aforementioned studies motivate our exploration of interactions of surface elastic waves propagating in a thick plate (with soil parameters) structured with concrete pillars above it. Pillars are $40$ m in height and the plate is $100$ m in thickness, so that typical frequencies under study are below $1$ Hz, a frequency range of particular interest in earthquake engineering. We demonstrate that three seismic cloaks allow for an  unprecedented flow of elastodynamic energy. These designs are achieved by first computing ideal cloaks' parameters deduced from a geometric transform in the Navier equations that leads to almost isotropic and symmetric elasticity (4th order) and density (2nd order) tensors. To do this we extend the theory of Non-Euclidean cloaking for light as proposed by the theoretical physicists Leonhardt and Tyc to mechanical waves [Science 323, 110-112 (2009)]. In a second step, ideal heterogeneous nearly isotropic cloak's parameters are approximated by averaging elastic properties of sets of pillars placed at the nodes of a bipolar coordinate grid, which is an essential ingredient in our Non-Euclidean cloaking theory for elastodynamic waves. Cloaking effects are studied for a clamped obstacle (reduction of the disturbance of the wave wavefront and its amplitude behind a clamped obstacle). Protection is achieved through reduction of the wave amplitude within the center of the cloak.These results represent a first step towards designs of Non-Euclidean seismic cloaks for surface (Rayleigh and Love) waves
propagating in semi-infinite elastic media structured with pillars.

\end{abstract}
%\label{firstpage}
\maketitle

\section{Introduction}
In this paper, we would like to present a novel approach for the design of seismic cloaks, which is based on the extension of Non-Euclidean cloaking theory
  for electromagnetic waves as proposed in the seminal paper by Leonhardt and Tyc \cite{leonhardt-tyc}, to mechanical waves.
%s proposed in the seminal paper by Milton, Briane and Willis (NJP 2006) which investigated form invariant governing equations in physics. 

Although our theory applies to the full elastodynamic waves equations in two- and three-dimensional settings, and thus extends the former works of \cite{milton2006,farhat2009,brunapl,norris2011,diatta2014} on transformational elastodynamics to a Non-Euclidean setting (i.e. from flat to curved spaces). However, for the sake of simplicity in the numerical computations and physical discussions which are focused on seismic waves, we shall restrict ourselves to elastic waves propagating in plates, rather than in semi-infinite or three-dimensional media. This means that we will consider flexural, rather than Rayleigh surface waves (but from a former work on clamped seismic metamaterials \cite{achaoui17a}, we know that study of the former provides invaluable information on the latter). For this case, we draw analogies between the transformed Helmholtz equation that governs propagation of transverse electromagnetic waves and the transformed Kirchhoff-Love equation that governs propagation of flexural waves, making use of results in \cite{tyc10}. We shall then consider the more challenging case of in-plane polarized Love-like waves, that requires a whole new Non-Euclidean elastodynamic theory of its own.

The full Non-Euclidean elastodynamic theory (NEET) might seem fairly involved and thus difficult to apply in practice, but as noted in \cite{prl2014,achaoui17a}, the elastic plate model already gives some interesting insight in the physics of seismic waves. This is the reason why we first present some derivation of transformed Kirchhoff-Love, which would be enough to get a grasp on the physics of surface Rayleigh waves propagating in Non-Euclidean transformed soils. The section on complete NEET for Navier equations with an almost isotropic elasticity tensor can be skipped by readers who wish to implement Non-Euclidean seismic cloaks for Rayleigh waves. However, understanding the case of Love-like waves requires reading through the NEET section. 

We also explain in this paper how one can approximate these spatially varying anisotropic elastic parameters with a homogenization approach.
Some finite element computations are given for the transformed elasticity system in its approximate structured media.
Comparisons between these numerical results demonstrate that one can achieve some control of flexural and coupled in-plane shear and pressure wave trajectories. Bearing in mind that we use soil and concrete elastic parameters for the
plate and pillars, and that we consider wave frequencies lower than $1$ Hz, we claim that our results represent a preliminary step towards implementation of Non-Euclidean seismic cloaks in civil engineering. Importantly, we point out that this Non-Euclidean setting considerably reduces the anisotropy and minor symmetry breaking of the elasticity tensor one gets using the Euclidean transformation elastodynamic theory \cite{brunapl,diatta2014}, which has important consequences for practical implementation of seismic cloaks: one need not look for structured elastic media displaying some form of chirality e.g. in the framework of micropolar elasticity \cite{achaoui17b}, or mimicking a simplified form of Willis media \cite{willis1981,diatta16a}.  

%\section{Transformed equation and elastic properties of cloak}
\section{Non-Euclidean transformation Cloaking in bipolar coordinates}

Elastic waves, in the same way as light, are subject to Snell's law of refraction, as it transpires from the seismic ray theory \cite{cerveny} and an appropriate choice of material properties leads to spectacular effects like those created by gradient index lenses and carpet cloaks in optics \cite{li08a}. Compared to a cloak designed with transformation optics \cite{pendry06} where extreme anisotropic parameters are required, within a conformal optics cloak \cite{leonhardt06} or a carpet cloak designed with a quasi-conformal grid \cite{li08a}, ray-paths are bent thanks to a smooth refractive index transition. Maxwell \cite{maxwell} studied gradient index lenses that manipulate wave trajectories in unconventional manner, notably Maxwell's fisheye whose spatially varying refractive index was later shown to be associated with a stereographic projection of a sphere on a plane \cite{luneburg,leonhardt09}. This was used in \cite{colombi16a}, wherein the concepts advanced by Luneburg translate nearly unchanged to a transformation seismology theory that requires approximating waves by seismic rays.

%({\bf Transcription de l'article d'Ulf Leonhardt et   Thomas Tyc pour les ondes de flexion, puis les ondes de pression/cisaillement ? })

In what follows, we propose to apply unexplored new models  of elastodynamic cloaks based on non-Euclidean cloaking, with applications to  seismic cloaking.
This is an adaptation to elastodynamics  of an earlier proposal by Ulf Leonhardt and Thomas Tyc \cite{leonhardt-tyc}  of an invisibility cloak 
with spatially varying refractive index with weak anisotropy.
To do this, we need to extend the mathematics of  non-Euclidean cloaking to elastodynamic waves. Bipolar coordinates are particularly well suited  for this study.
One starts from the Navier equations in isotropic elastic medium in bipolar coordinates and then applies the transformation techniques.

\subsection{The bipolar coordinates}\label{sect:bipolar}

To describe the cloaks, we use bipolar
cylindrical coordinates $(\tau,\sigma,z)$. They are related
to the Cartesian coordinates $(x, y, z)$ as
\begin{equation}
x=\frac{a\sinh\tau}{\cosh\tau-\cos\sigma} \; , \; y=\frac{a\sin\sigma}{\cosh\tau-\cos\sigma} \; , \; z=z \; .
\label{tyc1}
\end{equation}
In practice, the parameter $a$ stands for the size of the cloak. The position vector corresponding to a point with coordinates $(\sigma,\tau,z),$  is then
\begin{eqnarray}
{\bf r}= \frac{a\sinh\tau}{\cosh\tau-\cos\sigma} {\bf i}+ \frac{a\sin\sigma}{\cosh\tau-\cos\sigma}{\bf j}+z{\bf k}.
\end{eqnarray}

The Jacobian of the change of coordinates between Cartesian and bipolar coordinates is given by the following equalities
\begin{eqnarray}
\frac{\partial x }{\partial\tau} = - \frac{\partial y }{\partial\sigma}= a\frac{1-\cosh(\tau)\cos(\sigma)}{(\cosh(\tau)-\cos(\sigma))^2}; ~~~ \frac{\partial x }{\partial\sigma} = \frac{\partial y }{\partial\tau}=-a\frac{\sin(\sigma)\sinh(\tau)}{(\cosh(\tau)-\cos(\sigma))^2};
\end{eqnarray}
so that the corresponding  orthogonal basis is 
\begin{eqnarray}
{\bf e}_1:= {\bf e}_\sigma&=& \frac{\partial \bf r }{\partial\sigma}=-a\frac{\sin(\sigma)\sinh(\tau)}{(\cosh(\tau)-\cos(\sigma))^2}~ {\bf i} -a\frac{1-\cosh(\tau)\cos(\sigma)}{(\cosh(\tau)-\cos(\sigma))^2}~{\bf j}
\\
{\bf e}_2:= {\bf e}_\tau&=& \frac{\partial \bf r }{\partial\tau} =a\frac{1-\cosh(\tau)\cos(\sigma)}{(\cosh(\tau)-\cos(\sigma))^2}~{\bf i}-a\frac{\sin(\sigma)\sinh(\tau)}{(\cosh(\tau)-\cos(\sigma))^2}~ {\bf j} ;~~ 
\\
{\bf e}_3:={\bf e}_z&=& \frac{\partial \bf r }{\partial z} = {\bf k}.
\end{eqnarray}
As one can see,  ${\bf e}_3$ is a unit vector, whereas, ${\bf e}_1$ and ${\bf e}_2$ have equal norm 
$||\frac{\partial \bf r }{\partial\sigma}||=||\frac{\partial \bf r }{\partial\tau}||= \frac{a} {\cosh\tau-\cos\sigma}.$ 
We then consider the orthonormal basis 
\begin{eqnarray}
\hat{\bf e}_1:=\hat{\bf e}_\sigma&=&\frac{1}{||\frac{\partial \bf r }{\partial\sigma}||} \frac{\partial \bf r }{\partial\sigma}=
-\frac{\sin(\sigma)\sinh(\tau)}{\cosh\tau-\cos\sigma}~ {\bf i} -\frac{1-\cosh\tau\cos\sigma}{\cosh(\tau-\cos\sigma}~{\bf j}
\\
\hat{\bf e}_2:= \hat{\bf e}_\tau&=&\frac{1}{||\frac{\partial \bf r }{\partial\tau}||} \frac{\partial \bf r }{\partial\tau} =\frac{1-\cosh\tau\cos\sigma}{\cosh\tau-\cos\sigma}~{\bf i}-\frac{\sin\sigma\sinh\tau}{\cosh\tau-\cos\sigma}~ {\bf j} ;~~ 
\\
\hat{\bf e}_3:={\bf e_z}&=&  {\bf k}.
\end{eqnarray}
One can also notice the following identities
\begin{eqnarray}
\frac{\partial \hat{\bf e}_\sigma}{\partial \sigma}& =& \frac{\sinh\tau}{\cosh\tau-\cos\sigma }\hat{\bf e}_\tau, ~~~
\frac{\partial \hat{\bf e}_\sigma}{\partial \tau} =-\frac{\sin\sigma}{\cosh\tau-\cos\sigma}\displaystyle\hat{\bf e}_\tau
\nonumber
\\
\frac{\partial \hat{\bf e}_\tau}{\partial \sigma}& =&- \frac{\sinh\tau}{\cosh\tau-\cos\sigma }\hat{\bf e}_\sigma, ~~~
\frac{\partial \hat{\bf e}_\tau}{\partial \tau} =\frac{\sin\sigma}{\cosh\tau-\cos\sigma}\displaystyle\hat{\bf e}_\sigma.
\end{eqnarray}
One also further considers the inverse of the Jacobian of the change of coordinates between Cartesian and bipolar coordinates via the following 
\begin{eqnarray}
\frac{\partial \sigma }{\partial x}=- \frac{1}{a}\sin\sigma\sinh\tau, ~~\frac{\partial \sigma }{\partial y}=  \frac{1}{a}(1-\cosh\tau\cos\sigma),
~~
\frac{\partial \tau }{\partial x}=  \frac{1}{a}(\cosh\tau\cos\sigma-1), ~~  \frac{\partial \tau }{\partial y}= - \frac{1}{a}\sin\sigma\sinh\tau.
\end{eqnarray}

\subsection{A description of the transformation}
%To describe the Euclidean part of the cloak, we use bipolar cylindrical coordinates $(\tau,\sigma,z)$ in physical space that are related to the Cartesian coordinates $(x, y, z)$ as
%\begin{equation}x=\frac{a\sinh\tau}{\cosh\tau-\cos\sigma} \; , \; y=\frac{a\sin\sigma}{\cosh\tau-\cos\sigma} \; , \; z=z \; . \label{tyc1} \end{equation}
%Here, the parameter $a$ characterizes the size of the cloak. Analogous relations hold in virtual space where $x$, $y$, and $\sigma$ are replaced by $x'$, $y'$, and $\sigma'$, respectively. 

The mapping between
the two spaces is given in the bipolar cylindrical coordinates by
the following relation between $\sigma$ and $\sigma'$:
\begin{equation}
\sigma'=\sigma \; , \; \rm{for} \mid\sigma\mid\leq\pi/2 \; ,
\label{tyc2}
\end{equation}

and

\begin{equation}
\sigma'=\left(\frac{4\sigma^2}{\pi} - 3\mid\sigma\mid + \pi \right) \; , \; \rm{for} \frac{\pi}{2}<\mid\sigma\mid<\frac{3\pi}{4} \; ,
\label{tyc3}
\end{equation}
and the coordinates $\tau$, $z$ coincide in the two spaces (the mapping only involves inplane coordinates). The line
element $ds^2 = d{x'}^2 +d{y'}^2 +d{z'}^2$ in virtual space is equal to
\begin{equation}
ds^2=\frac{a^2}{{(\cosh\tau-\cos\sigma')}^2} (d\tau^2+d{\sigma'}^2)+dz^2 \; .
\label{tyc4}
\end{equation}

%{\bf ( Seb, c'est la meme transformation que tu vas decrire pour les "flexural waves". Il serait bien de mettre cette destription ici et l'utiliser dans les sections suivantes.)}

\begin{figure}[h!]
\resizebox{70mm}{!}{\includegraphics{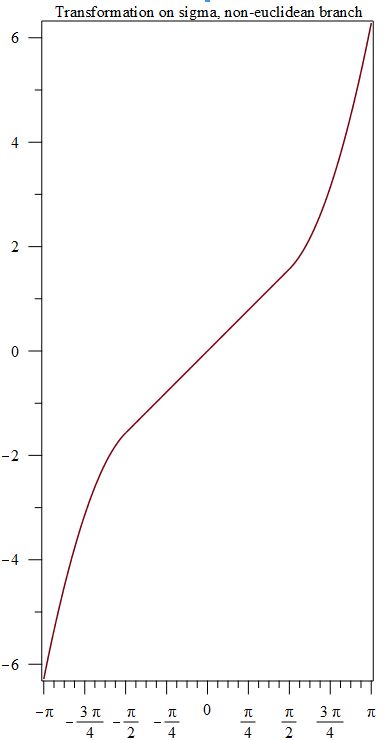}}
\vspace{-3mm}
\caption{Transformation $\sigma\mapsto \sigma'$, with $\sigma'=\sigma$ if $|\sigma| \leq \frac{\pi}{2}$ and   $\sigma'=\text{sgn}(\sigma) \Big(\frac{4\sigma^2}{\pi}-3|\sigma|+\pi\Big) $ when  $ \frac{\pi}{2} <|\sigma| \leq  \frac{3\pi}{4}$
}
\label{transformation}
\end{figure}

The mapping between the non-Euclidean part of virtual space
and the corresponding part of physical space requires more work.
Here, we will not reproduce all the corresponding formulas of
\cite{tyc09b}, but give some important steps, as was done in \cite{tyc10}. One
starts with a sphere of radius $r=(4/\pi) a$
(see Eq. (S43) in \cite{tyc09b}) parameterized by spherical coordinates—
the latitudinal angle $\theta$ and the longitudinal angle that is denoted
by $\sigma'$. This is not yet the sphere of virtual space, but is related
to it by a suitable M\" obius transformation (see Eq. (S40) in \cite{tyc09b}) represented
on the sphere via stereographic projection (see Eq. (S37)-(S39) in \cite{tyc09b}).
Then, the coordinates $(\sigma',\theta)$ are mapped to the
bipolar coordinates $(\sigma,\tau)$ of physical space as follows. The
mapping between $\sigma'$ and $\sigma$ is given by (\ref{tyc3}),
where $\sigma'$ is set to
one of the intervals $[-2\pi,-\pi]$ or $[\pi, 2\pi]$ by adding or subtracting
$2\pi$. The corresponding intervals of $\sigma$ are then $[-\pi,-3\pi/4]$
and $[3\pi/4, \pi]$, see Fig. \ref{transformation}. The mapping between $\theta$ and $\tau$ can be derived by
combining Eqs. (S41) and (S46) in \cite{tyc09b}. After some calculations,
this yields
\begin{equation}
\theta=2\arctan(t-1+\sqrt{t^2+1} \; , \; \rm{ with } \; t=\tan\left( \frac{\pi}{4} \left( \frac{\sinh\tau}{\cosh\tau+1}\right) + \frac{\pi}{4} \right) \; .
\label{tyc7}
\end{equation}

The line element of the non-Euclidean part of virtual space is

\begin{equation}
ds^2= \frac{16a^2}{\pi^2} {\left( \frac{1+\cot^2(\theta/2)}{1-2\cos\sigma'\cot(\theta/2)+2\cot^2(\theta/2)}\right)}^2 (d\theta^2+\sin^2\theta d{\sigma'}^2)+dz^2 \; .
\label{tyc8}
\end{equation}

This line element is slightly different from the line element in Eq. (S47) of \cite{tyc09b}, as noted in \cite{tyc10}, where details can be found.

\subsection{Non-Euclidean cloaking for flexural waves}
Let us now apply the aforementioned transformation to the case of flexural waves. For this, we make use of the correspondence between the transformed anisotropic biharmonic equation
for flexural waves and the anisotropic equation for transverse
electromagnetic waves \cite{farhat2009} to extend the concept of non-Euclidean cloaking to plates:
\begin{equation}
\mu^{-1}\nabla\cdot[\underline{\underline{\varepsilon}}^{-1}\nabla H_z(x,y)]+ \frac{\omega^2}{c^2} H_z(x,y)=0
\label{max}
\end{equation}
where $\underline{\underline{\varepsilon}}$ is a tensor of relative permittivity, $\mu$ is the isotropic relative permeability,
$c$ the speed of light in vacuum and $\omega$ the angular wave frequency.

We perform the geometric transformation from
2D virtual space to the plane $xy$ of physical space, assuming that the material parameters
of the plate only depend upon $xy$ coordinates, thereby generating
a thin 3D physical space, with small plate thickness in z direction.
For calculating the line elements in this 3D physical space, we then
have to add the term $dz^2$ to the line elements corresponding to
the 2D situation.

As is well known, the out of plane displacement ${\bf u}=(0,0,u(x,y))$ in the $z$ direction is
solution of the Kirchhoff-Love equation
\begin{equation}
\rho^{-1}\nabla\cdot[\underline{\underline{\xi}}^{-1}\nabla(\rho^{-1}\nabla\cdot[\underline{\underline{\xi}}^{-1}\nabla u(x,y)]-\beta^4 u(x,y)=0
\label{love}
\end{equation}
where $\rho$ is the plate density inside the cloak, $\xi$ its anisotropy Young modulus and
$\beta^4=\omega^2\rho_0 h/D_0$ is related to the angular wave frequency $\omega$, the plate
density $\rho_0$ (outside the cloak), its flexural rigidity $D_0$
and finally the constant plate thickness $h$.

The anisotropic Young modulus for this region is equal to the tensors of permittivity and permeability which have been in fact calculated by Tyc et al. in \cite{tyc10} in the case of the transverse electromagnetic waves governed by
the Helmholtz equation, using the same general method
as in \cite{leonhardt-tyc}, see also the supplemental material \cite{tyc09b}. A short, but
insightful, review of \cite{leonhardt-tyc} has been written by experts in generalized
cloaking \cite{nicolet09a} and can serve as an entry door in this field. In
\cite{tyc10}, the following expressions can be found:
\begin{equation}
\underline{\underline{\varepsilon}}
=\underline{\underline{\mu}}
=\rm{Diag} \left( \frac{d\sigma}{d\sigma'},\frac{d\sigma'}{d\sigma},\frac{(\cosh\tau-\cos\sigma)}{(\cosh\tau-\cos\sigma')}\frac{d\sigma'}{d\sigma}\right) \; ,
\label{tyc5}
\end{equation}
which can be translated in terms of anisotropic Young modulus and density in the anisotropic Kirchhoff-Love Eq. (\ref{love}):
\begin{equation}
\underline{\underline{\xi}}=\rm{Diag} \left( \frac{d\sigma}{d\sigma'},\frac{d\sigma'}{d\sigma} \right)
\rm{ and } \; \rho =\frac{(\cosh\tau-\cos\sigma)}{(\cosh\tau-\cos\sigma')}\frac{d\sigma'}{d\sigma} \; .
\label{tyc6}
\end{equation}

Tyc et al. simplify Eq. (\ref{tyc8}) in a more amenable form in \cite{tyc10} that leads to the following entries for the tensors of permittivity and permeability
\begin{equation}
\begin{array}{ll}
\underline{\underline{\varepsilon}} & =\underline{\underline{\mu}}= \rm{Diag} \left( \varepsilon_\sigma, \varepsilon_\tau, \varepsilon_z \right) \; ,
\hbox{ with } \\
%\label{tyc9}
%\end{equation}
%
%\begin{equation}
%\begin{array}{ll}
\varepsilon_\sigma & = \displaystyle{\frac{\pi}{4}\frac{(t^2+1+t\sqrt{t^2+1})}{{({t-1+\sqrt{t^2+1}})}} \frac{1}{(\cosh\tau+1)}\frac{d\sigma}{d\sigma'}} \\
\varepsilon_\tau &= \displaystyle{\frac{1}{\varepsilon_\sigma}}\\
\varepsilon_z &= \displaystyle{\frac{16}{\pi}\frac{(t-1+\sqrt{t^2+1}) (t^2+1+t\sqrt{t^2+1})}{{(1+{(t-1+\sqrt{t^2+1})}^2)}^2}
\frac{u^2 {(\cosh\tau-\cos\sigma)}^2}{(\cosh\tau+1)}\frac{d\sigma'}{d\sigma}} \; .
\end{array}
\label{tyc10}
\end{equation}

Regarding the kirchhoff-Love equation, this translates as follows in terms of anisotropic Young modulus and isotropic density
\begin{equation}
\underline{\underline{\xi}}=\rm{Diag} \left( \varepsilon_\sigma, \varepsilon_\tau \right)
\rm{ and } \; \rho= \varepsilon_z \; .
\label{tyc6}
\end{equation}

\subsection{Navier Equations in bipolar coordinates}

We will assume a  time
harmonic  $exp(i\omega t)$  dependence of the propagation of elastic waves, so that,  suppressing such a dependence, the governing Navier equations  in Cartesian coordinates
are 
\begin{eqnarray}
\frac{\partial}{{\partial x'_i}}\left({C'}_{ijkl}\frac{\partial}{{\partial x'_k}}{u'}_l\right)+{\rho'}_{jl}\omega^2 {u'}_{l}=0 \;   ,
%\nabla\cdot \left[ \C :\nabla  {\bf  u}
% \right]  + \rho\omega^2{\bf  u}=0
\label{nav1}
\end{eqnarray}
where the body force is assumed to be zero,  with $\omega$ the  angular  wave  frequency, $t$ the  time variable,
${\C}=(C_{ijkl}) $   the 4-order elasticity tensor and $\rho$  the density.

Let us reconsider the bases $({\bf e_1},{\bf e_2},{\bf e_3}) =  ({\bf e}_\sigma, ~{\bf e}_\tau, ~{\bf k})$ and  $ (\hat{\bf e_1}, ~\hat{\bf e_2}, ~\hat{\bf e_3})= (\hat{\bf e}_\sigma, ~\hat{\bf e}_\tau, ~{\bf k})$  specified in Section \ref{sect:bipolar}.
The corresponding metric (first fundamental form) is given by its coefficients $g_{ij}={\bf e_i}\cdot {\bf e_j},$ $i,j=1,2,3$ with 
\begin{eqnarray}
g_{11}= {\bf e_1}\cdot {\bf e_1}= \frac{a^2}{{(\cosh\tau-\cos\sigma')^2}} ,  ~~~g_{22}= {\bf e_2}\cdot {\bf e_2}=\frac{a^2}{{(\cosh\tau-\cos\sigma')^2}}, ~~ ~g_{33}=1. ~~ ~g_{ij}=0,\text{ if } i\neq j.
\end{eqnarray}

The expression of the gradient of a function $f$ in bipolar coordinates reads
\begin{eqnarray}
\nabla f =g^{ij}\partial_if ~\frac{\bf e_j}{||{\bf e_j}||} = \frac{ g^{ij}}{\sqrt{ g_{jj}}} \partial_if ~\hat {\bf e_j} =\frac{\cosh\tau-\cos\sigma}{a}\Big(\hat{\bf e}_\sigma ~\frac{\partial f }{\partial \sigma} 
+
\hat{\bf e}_\tau ~\frac{\partial f }{\partial \tau} \Big) +
\frac{\partial f }{\partial z} ~{\bf k}
%
%h_1 \frac{\partial f}{\partial \sigma} {\bf e_\sigma}
%+h_2 \frac{\partial f}{\partial \tau} {\bf e_\tau}+ h_3 \frac{\partial f}{\partial z} {\bf e_z}, ~~~  h_1= h_2=  \frac{a}{{(\cosh\tau-\cos\sigma')}}, ~~~ h_3=1,
% \right]  + \rho\omega^2{\bf  u}=0
%\label{nav1}
\end{eqnarray}
so we can write in a more general way
\begin{eqnarray}
\nabla
 &=&
{\hat{\bf e}_\sigma}~\frac{\cosh\tau-\cos\sigma}{a} ~\frac{\partial }{\partial \sigma} 
+
{\hat{\bf e}_\tau}~\frac{\cosh\tau-\cos\sigma}{a}  ~\frac{\partial  }{\partial \tau}  +
 ~{\bf k} \frac{\partial  }{\partial z}
\nonumber
\end{eqnarray}

In the basis $\Big(\hat{\bf e}_\sigma, \hat{\bf e}_\tau, {\bf k}\Big)$, the gradient of a vector field ${\bf u} = u_\sigma\hat{\bf e}_\sigma+ u_\sigma\hat{\bf e}_\tau+ u_z{\bf k}$,  is represented by the following matrix
\begin{eqnarray}
\nabla {\bf u}=\begin{pmatrix}\frac{\cosh\tau-\cos\sigma}{a} \frac{\partial u_\sigma}{\partial \sigma}-
\frac{\sinh\tau}{a}
 u_\tau & 
\frac{\cosh\tau-\cos\sigma}{a}\frac{\partial u_\tau}{\partial \sigma} +
\frac{\sinh\tau}{a}
 u_\sigma &   \frac{\partial u_z}{\partial \sigma}
\\ 
\frac{\cosh\tau-\cos\sigma}{a} \frac{\partial u_\sigma}{\partial \tau}+
\frac{\sinh\tau}{a}u_\tau & 
\frac{\cosh\tau-\cos\sigma}{a} \frac{\partial u_\tau}{\partial \tau}
- \frac{\sinh\tau}{a}u_\sigma
& \frac{\partial u_z}{\partial \tau} 
\\
\frac{\partial u_\sigma}{\partial z}  &  \frac{\partial u_\tau}{\partial z}  &   \frac{\partial u_z}{\partial z} 
\end{pmatrix}
\label{nablabipolar}
\end{eqnarray}
We are now ready to write the Navier Equation in bipolar coordinates.   It reads

\begin{eqnarray} 
- \sqrt{-1}\omega p_\sigma &=&   \frac{\cosh\tau-\cos\sigma}{a} \frac{\partial T_{\sigma\sigma} }{\partial \sigma} - \frac{\sinh\tau}{a}\Big(T_{\sigma\tau}+T_{\tau\sigma}\Big) 
%\nonumber\\%&+&\Big[
+\frac{\cosh\tau-\cos\sigma}{a}~ \frac{\partial T_{\tau\sigma} }{\partial \tau} - \frac{\sin\sigma}{a }\Big(T_{\sigma\sigma}-T_{\tau\tau}\Big) 
%\Big] \hat{\bf e}_\sigma
+ \frac{\partial T_{z\sigma} }{\partial \sigma} 
\nonumber\\
- \sqrt{-1}\omega p_\tau &=&
  \frac{\cosh\tau-\cos\sigma}{a} \frac{\partial T_{\sigma\tau} }{\partial \sigma} + \frac{\sinh\tau}{a }\Big(T_{\sigma\sigma}-T_{\tau\tau}\Big) 
+  \frac{\cosh\tau-\cos\sigma}{a}~\frac{\partial T_{\tau\tau} }{\partial \tau} - \frac{\sin\sigma}{a }\Big(T_{\sigma\tau}+T_{\tau\sigma}\Big) 
 +\frac{\partial T_{z\tau} }{\partial \tau}
\nonumber\\
- \sqrt{-1}\omega p_z &=& \frac{\cosh\tau-\cos\sigma}{a}~ \frac{\partial T_{\sigma z} }{\partial \sigma} - \frac{\sinh\tau}{a }T_{\tau z} 
+\frac{\cosh\tau-\cos\sigma}{a}~ \frac{\partial T_{\tau z} }{\partial \tau} - \frac{\sin\sigma}{a }T_{\sigma z} +
\frac{\partial T_{zz} }{\partial z}
\label{NavierBipolar}
\end{eqnarray}
where ${\bf T:=\mathbb C:\nabla {\bf u}}$ is the stress tensor in bipolar coordinates, with components $T_{ij}= \mathbb C_{ijkl}\Big(\nabla {\bf u}\Big)_{kl}$  and $\nabla {\bf u}$  is as in (\ref{nablabipolar}).
Here we have relabeled the indices $(\sigma, \tau,z)$ by $(1,2,3)$, in the components $\mathbb C_{ijkl}$ of the tensor $\mathbb C.$

\subsection{Elastodynamic non-Euclidean cloaking}
The Navier Equations (\ref{NavierBipolar}) retain their form under any transformation $(\sigma,\tau, z) \mapsto (\sigma',\tau', z')$ and now read
\begin{eqnarray} 
- \sqrt{-1}\omega ~p_{\sigma'} &=&   \frac{\cosh\tau'-\cos\sigma'}{a} \frac{\partial T_{\sigma'\sigma'}' }{\partial \sigma'} - \frac{\sinh\tau'}{a}\Big(T_{\sigma'\tau'}'+T_{\tau'\sigma'}'\Big) 
%\nonumber\\%&+&\Big[
+\frac{\cosh\tau'-\cos\sigma'}{a}~ \frac{\partial T_{\tau'\sigma'}' }{\partial \tau'} - \frac{\sin\sigma'}{a }\Big(T_{\sigma'\sigma'}'-T_{\tau'\tau'}'\Big) 
%\Big] \hat{\bf e}_\sigma
+ \frac{\partial T_{z'\sigma'}' }{\partial \sigma'} 
\nonumber\\
- \sqrt{-1}\omega ~p_{\tau'} &=&
  \frac{\cosh\tau'-\cos\sigma'}{a} \frac{\partial T_{\sigma'\tau'}' }{\partial \sigma'} + \frac{\sinh\tau'}{a }\Big(T_{\sigma'\sigma'}'-T_{\tau'\tau'}'\Big) 
+  \frac{\cosh\tau'-\cos\sigma'}{a}~\frac{\partial T_{\tau'\tau'}' }{\partial \tau'} - \frac{\sin\sigma'}{a }\Big(T_{\sigma'\tau'}'+T_{\tau'\sigma'}'\Big) 
 +\frac{\partial T_{z'\tau'}' }{\partial \tau'}
\nonumber\\
- \sqrt{-1}\omega ~p_{z'} &=& \frac{\cosh\tau'-\cos\sigma'}{a}~ \frac{\partial T_{\sigma' z'}' }{\partial \sigma'} - \frac{\sinh\tau'}{a }T_{\tau' z'}' 
+\frac{\cosh\tau'-\cos\sigma'}{a}~ \frac{\partial T_{\tau' z'} '}{\partial \tau'} - \frac{\sin\sigma'}{a }T_{\sigma' z'}' +
\frac{\partial T_{z'z'}'}{\partial z'}
\label{NavierBipolarTransformed}
\end{eqnarray}
where ${\bf T':=\mathbb C':\nabla {\bf u'}}$ is the transformed stress tensor in bipolar coordinates, with components $T_{ij}'= \mathbb C_{ijkl}'\Big(\nabla' {\bf u}\Big)_{kl}.$  As above, $\nabla' {\bf u}$ is the gradient of  the displacement ${\bf u}$, now expressed in the transformed bipolar coordinates.

We are interested in the situation where the displacement ${\bf u}$ only depends upon the  variables  $\sigma, \tau$ and not on $z.$  
In this case, the above gradient (\ref{nablabipolar}) reads
\begin{eqnarray}
\nabla {\bf u}=\begin{pmatrix}\frac{\cosh\tau-\cos\sigma}{a} \frac{\partial u_\sigma}{\partial \sigma}-
\frac{\sinh\tau}{a}
 u_\tau & 
\frac{\cosh\tau-\cos\sigma}{a}\frac{\partial u_\tau}{\partial \sigma} +
\frac{\sinh\tau}{a}
 u_\sigma &   \frac{\partial u_z}{\partial \sigma}
\\ 
\frac{\cosh\tau-\cos\sigma}{a} \frac{\partial u_\sigma}{\partial \tau}+
\frac{\sinh\tau}{a}u_\tau & 
\frac{\cosh\tau-\cos\sigma}{a} \frac{\partial u_\tau}{\partial \tau}
- \frac{\sinh\tau}{a}u_\sigma
& \frac{\partial u_z}{\partial \tau} 
\\
0  &  0 &  0
\end{pmatrix} \; .
\end{eqnarray}

We also consider transformations of the form $(\sigma,\tau,z) \mapsto \Big(\sigma'(\sigma,\tau),\tau'(\sigma,\tau),z'(z)\Big)$ in bipolar coordinates, with Jacobian 
\begin{eqnarray}J=J_{\sigma'\sigma}&=&
\begin{pmatrix}\frac{\partial\sigma'}{\partial \sigma}
 & 
\frac{\partial\tau'}{\partial \sigma} &   0
\\ 
 \frac{\partial \sigma'}{\partial \tau} & 
 \frac{\partial \tau'}{\partial \tau}
& 0
\\
0  &  0  &   \frac{\partial z'}{\partial z} 
\end{pmatrix} .
\end{eqnarray}
More precisely, we are interested in transformations in $(\sigma,\tau)-$plane. In this case, the transformed elasticity tensor $\mathbb C'$ has the following components
\begin{eqnarray}
\mathbb C'_{ijkl} &=&  \det(J_{\sigma'\sigma}^{-1})  ~\frac{\cosh\tau'-\cos\sigma'}{\cosh\tau-\cos\sigma}~ 
 J_{ip}~ J_{kn}~ \mathbb C_{pjnl}, ~~ \text{ for } i,j,k,l=1,2. 
\end{eqnarray}

In this paper, we specialize to transformations of the form $(\sigma,\tau,z) \mapsto (\sigma'(\sigma),\tau,z)$ so that the Jacobian is of the form 
$J_{\sigma'\sigma}=\begin{pmatrix} \frac{\partial \sigma'}{\partial \sigma} &0& 0\\
0 &1& 0\\ 
0& 0& 1\end{pmatrix}.$
In particular, the nonvanishing coefficients of the transformed elasticity tensor
$\mathbb C'$ are explicitly given as 
\begin{eqnarray}
\mathbb C'_{\sigma \sigma\sigma\sigma} &=&  \frac{\partial \sigma'}{\partial \sigma}   ~\frac{\cosh\tau-\cos\sigma'}{\cosh\tau-\cos\sigma}~ (\lambda+2\mu),
\nonumber
\\
\mathbb C'_{\sigma\sigma\tau\tau} &=&     ~\frac{\cosh\tau-\cos\sigma'}{\cosh\tau-\cos\sigma}~ 
\lambda, 
\nonumber
\\
\mathbb C'_{\sigma\tau\sigma\tau} &=&   \frac{\partial \sigma'}{\partial \sigma}  ~\frac{\cosh\tau-\cos\sigma'}{\cosh\tau-\cos\sigma}~ \mu,
 \nonumber
\\
\mathbb C'_{\sigma\tau\tau\sigma} &=& \frac{\cosh\tau-\cos\sigma'}{\cosh\tau-\cos\sigma}~ 
\mu, 
\nonumber\\
\mathbb C'_{\tau\sigma\sigma\tau} &=&  \frac{\cosh\tau-\cos\sigma'}{\cosh\tau-\cos\sigma}~ 
 \mu\; ,  
\nonumber
\\
\mathbb C'_{\tau \sigma \tau \sigma} &=&  \Big(\frac{\partial \sigma'}{\partial \sigma}\Big)^{-1}  ~\frac{\cosh\tau-\cos\sigma'}{\cosh\tau-\cos\sigma}~ 
 \mu,
 \nonumber
\\
\mathbb C'_{\tau\tau\sigma\sigma} &=&   ~\frac{\cosh\tau-\cos\sigma'}{\cosh\tau-\cos\sigma}~ 
\lambda, 
\nonumber
\\
\mathbb C'_{\tau\tau\tau\tau} &=&  \Big(\frac{\partial \sigma'}{\partial \sigma}\Big)^{-1} ~\frac{\cosh\tau-\cos\sigma'}{\cosh\tau-\cos\sigma}~ 
 ~ (\lambda+2\mu)\; .
\end{eqnarray}

\section{Numerical results for pillar-based platonic non-Euclidean cloaks}
In this section, we would like to propose an approximate model of Non-Euclidean platonic cloak using some specific arrangement of pillars lying atop soft elastic plates, which have physical characteristics compatible (in bulk modulus and density) with concrete and soil, respectively. More precisely, we consider three types of cloaks consisting of $N_1$ (cloak 1), $N_2$ (cloak 2) and $N_3$ (cloak 3) pillars of height $40$ m atop a plate of thickness $100$ m. The cloak's inner and outer radii are $R_1=150$ m and $R_2=550$ m.
%(normalized with respect to the plate thickness).
The cloaks have pillars of same height, but of varying cross section (cross sectional area increases with distance from the center of each cloak).

\begin{figure}[h!]
\resizebox{120mm}{!}{\includegraphics{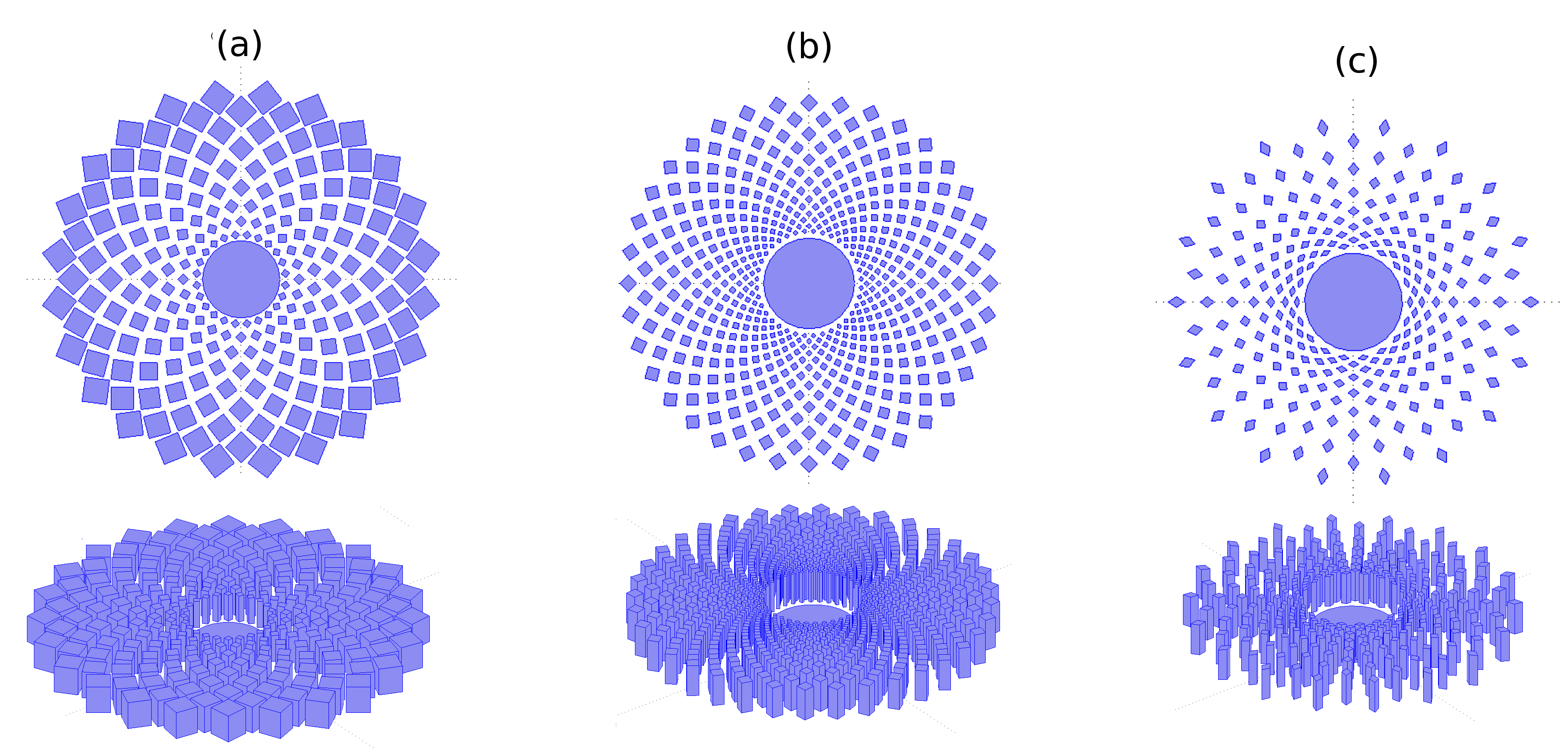}}
%{\includegraphics{fig1omega2to4nu01spherealone1.jpg}}
\vspace{-3mm}
\caption{Schematics of three cloaks  consisting of (a) $N_1=$ (cloak 1), (b) $N_2$ (cloak 2) and (c) $N_3$ (cloak 3) pillars of height $40$ m (in purple/gray color) atop a plate of thickness $100$ m surrounding a cylindrical obstacle of radius $130$ m (in purple/gray color). Upper panel is a top view and lower panel a 3D view. 
}
\label{cloaks_geometry}
\end{figure}

\begin{figure}[h!]
\resizebox{120mm}{!}{\includegraphics{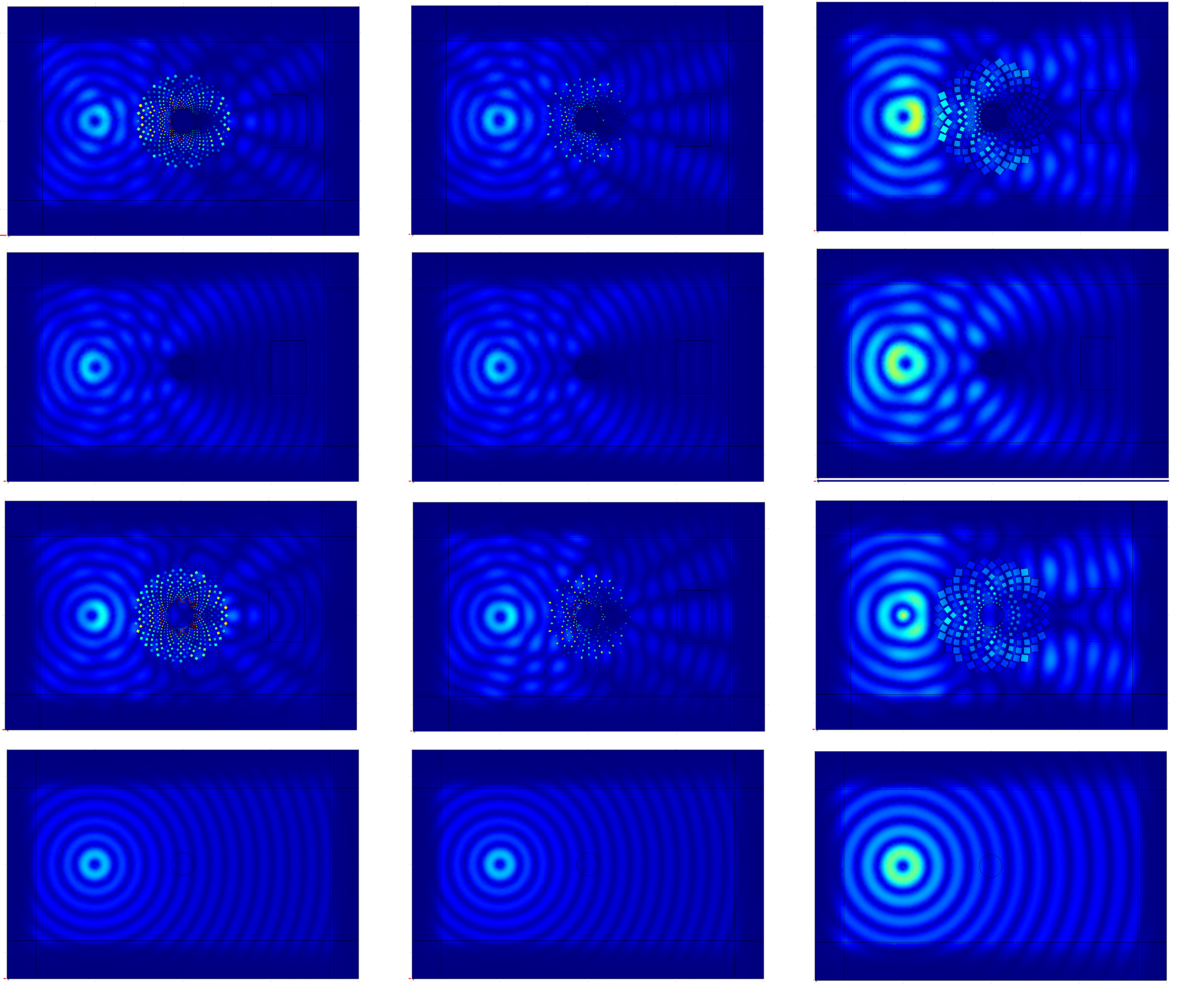}}
%{\includegraphics{fig1omega2to4nu01spherealone1.jpg}}
\vspace{-3mm}
\caption{Total displacement $\sqrt{u_1^2+u_2^2+u_3^2}$ for an out-of-plane point force (vertical polarization)
vibrating at frequency $x$Hz for a clamped obstacle with a cloak (1st row), without cloak (2nd row), and for soil
surrounded by a cloak (3rd row) by comparison with homogeneous soil (4th row). Left column is for cloak 1 as in Figure \ref{cloaks_geometry}, middle column for cloak 2, and right column for cloak 3.
}
\label{younes2}
\end{figure}

Numerical simulations are performed using the commercial finite element package Comsol Multiphysics, with specially designed perfectly matched layers described in \cite{diatta16a}. In figure \ref{younes2} (left panel), one can see that the cloak 1 displays a cloaking interval for flexural waves between $0.15$ Hz and $0.17$ Hz and besides from that the region in the center of the cloak is protected from $0.15$ Hz to $0.2$ Hz. However, the same cloak only works as an in-plane shear-wave protection from $0.2$ Hz to $0.25$ Hz, with no invisibility. Moreover, one can see some resonance peak at $0.28$ Hz, which can be attributed to the resonance of the stress-free obstacle.

In figure \ref{younes2} (middle panel), one can see that the cloak 2 displays a cloaking interval for flexural waves between $0.1$ Hz and $0.15$ Hz and besides from that the region in the center of the cloak is protected above $0.15$ Hz except for a resonance peak around $0.18$ Hz. However, the same cloak only works as an in-plane shear-wave protection from $0.18$ Hz to $0.36$ Hz, with no invisibility.

 In figure \ref{younes2} (right panel), one can see that the cloak 3 displays a cloaking interval for flexural waves between $0.08$ Hz and $0.14$ Hz and besides from that the region in the center of the cloak is protected above $0.18$ Hz except for a resonance peak around $0.23$ Hz. However, the same cloak only works as an in-plane shear-wave protection from $0.22$ Hz to $0.45$ Hz, with no invisibility.   

\begin{figure}[h!]
\resizebox{120mm}{!}{\includegraphics{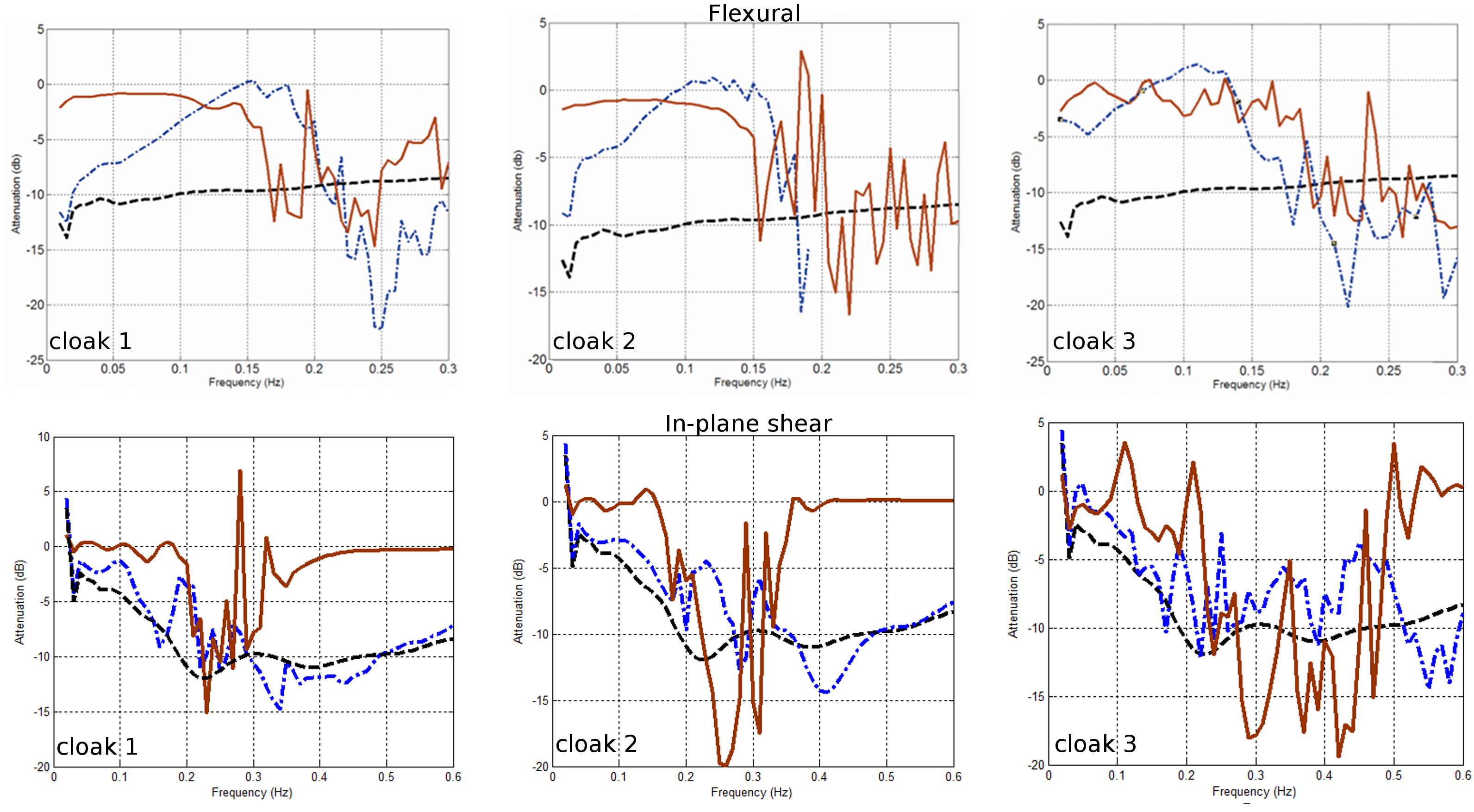}}
%{\includegraphics{fig1omega2to4nu01spherealone1.jpg}}
\vspace{-3mm}
\caption{Transmission loss retrieved behind (dotted blue curve) and inside (solid red curve) a cloak together with transmission loss induced by a clamped obstacle (dashed black curve) for three designs of concrete pillars of $40$ m
in height atop a plate of thickness $100$m with soil parameter;
Upper (resp. lower) panel describes the case of out-of-plane flexural (resp. in plane shear) waves.
}
\label{younes1}
\end{figure}

\begin{figure}[h!]
\resizebox{120mm}{!}{\includegraphics{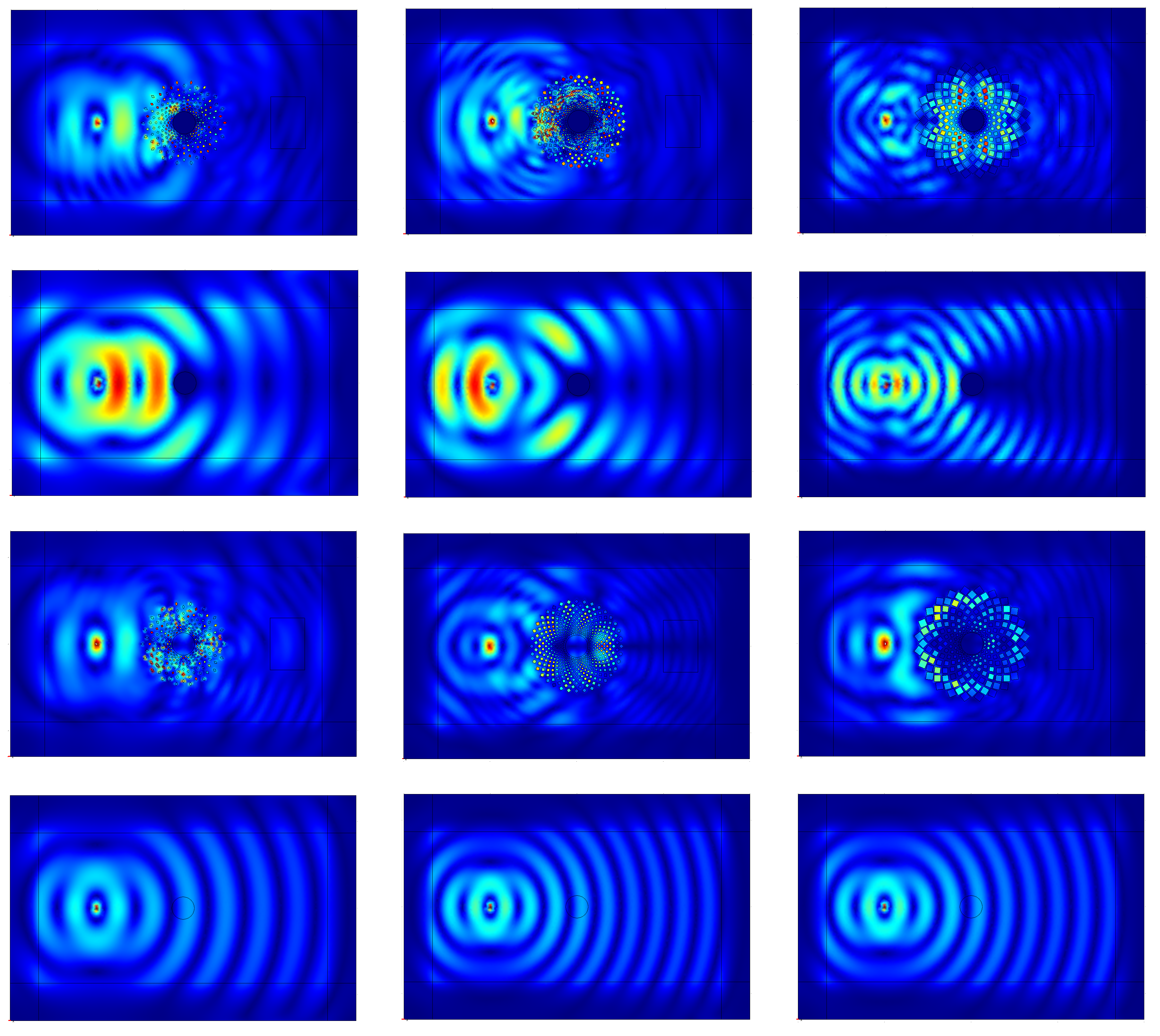}}
%{\includegraphics{fig1omega2to4nu01spherealone1.jpg}}
\vspace{-3mm}
\caption{Total displacement $\sqrt{u_1^2+u_2^2+u_3^2}$ for an in-plane shear polarized point force (vertical polarization) vibrating at frequency $x$Hz for a clamped obstacle with a cloak , and for soil
surrounded by a cloak (3rd row) by comparison with homogeneous soil (4th row). Left column is for cloak 1 as in
Figure \ref{cloaks_geometry}, middle column for cloak 2, and right column for cloak 3.
\label{younes3}
}
\end{figure}

Let us now look at the displacement field scattered by the three types of cloaks, at certain specific source frequencies
and out-of-plane (flexural) polarization. One can see in figure \ref{younes3} that the displacement field is enhanced within the cloak
(pillars behave like local resonators) and that the forward scattering of the flexural wave by a clamped obstacle is reduced in phase and magnitude once it has been surrounded by cloaks 1, 2 and 3.  

Let us now look at the displacement field scattered by the three types of cloaks, at certain specific source frequencies
and in-plane (shear) polarization. One can see in figure \ref{younes3} that the displacement field is enhanced within the cloak (pillars behave like local resonators) and that the forward scattering of the shear wave by a clamped obstacle is comparable in phase and magnitude with and without cloaks 1, 2 and 3. Hence, one might say that these cloaks fail
to offer interesting invisibility features. However, they are extremely efficient in terms of protection. It seems thus
fair to say that applications of cloaking to seismic wave protection goes beyond the invisibility principle. We believe
that inertia and dissipation produced by the pillars are instrumental in the good protection features. Such results
display marked differences with earlier attempts at platonic cloaking in flat Euclidean spaces \cite{stenger2012,farhat2012,misseroni}.

\section{Concluding remarks}
In this article, we have introduced a non-Euclidean cloaking theory for elastodynamic waves propagating in plates with out-of-plane (flexural) and in-plane (coupled shear and pressure) polarizations.
The former (flexural) waves are governed by a transformed Kirchhoff-Love equation that can be deduced from an earlier work by Tyc et al. \cite{tyc10} on non-euclidean cloaking for transverse electromagnetic waves
governed by the transformed Helmholtz equation. However, the latter in-plane elastic waves are goverened by a transformed Navier equation that requires a whole new theory presented here in a 2D setting, but
which can easily be generalized to a 3D setting. We have conclusively shown that concrete pillars judiciously designed (according to a bipolar coordinates lattice) atop a soft elastic plate
(with soil parameters), allows one to considerably reduce the shadow region behind a clamped and a stress-free
obstacle, and to further reduce the elastic field vibrations of the stress-free obstacle. The stress-free obstacle
is $130$ m in radius and $40$ m in depth, so can be used in first approximation as a model of building's foundation.
We note that our non-Euclidean seismic cloaks have a fairly similar cloaking interval than the cloak designed via homogenization in a multiply perforated plate in \cite{farhat2012},
but the experimental demonstration of platonic cloaking by Stenger et al. \cite{stenger2012} displays a larger cloaking interval. Therefore, we believe there is room for
improvement of our cloaks' s design.
We would like to add that our results could be translated into geophysics upon use of time domain computations,
for instance with Specfem software. It would be also important to study the effect of viscoelasticity on the
elastic wave propagation, which can be done with Comsol Multiphysics.

The authors acknowledge European funding through ERC Starting Grant ANAMORPHISM.

\end{document}